\documentclass[twoside]{article}
\usepackage{amsmath,amssymb,amsthm}
\usepackage{float}
\usepackage{placeins}
\usepackage{braket}
\usepackage{subcaption}
\usepackage{qic,epsfig}
\usepackage{graphicx}
\usepackage{placeins}
\usepackage{subfiles}

\begin{document}
\setlength{\textheight}{8.0truein}    

\runninghead{A new type of quantum walks based on decomposing quantum states}
            {C. Kiumi}

\normalsize\textlineskip
\thispagestyle{empty}
\setcounter{page}{1}

\alphfootnote

\fpage{1}

\centerline{\bf
A NEW TYPE OF QUANTUM WALKS 
}
\vspace*{0.035truein}
\centerline{\bf BASED ON DECOMPOSING QUANTUM STATES}
\vspace*{0.37truein}
\centerline{\footnotesize
Chusei Kiumi\footnote{
E-mail: kiumi-chusei-bf@ynu.jp}}
\vspace*{0.015truein}
\centerline{\footnotesize\it Graduate School of Engineering Science, Yokohama National University}
\baselineskip=10pt
\centerline{\footnotesize\it 79-5 Tokiwadai, Hodogaya, Yokohama, 240-8501, Japan}
\vspace*{0.225truein}

\abstracts{
In this paper, the 2-state decomposed-type quantum walk (DQW) on a line is introduced as an extension of the 2-state quantum walk (QW). The time evolution of the DQW is defined with two different matrices, one is assigned to a real component, and the other is assigned to an imaginary component of the quantum state. Unlike the ordinary 2-state QWs, localization and the spreading phenomenon can coincide in DQWs. Additionally, a DQW can always be converted to the corresponding 4-state QW with identical probability measures. In other words, a class of 4-state QWs can be realized by DQWs with 2 states. In this work, we reveal that there is a 2-state DQW corresponding to the 4-state Grover walk. Then, we derive the weak limit theorem of the class of DQWs corresponding to 4-state QWs which can be regarded as the generalized Grover walks.
}{}{}

\vspace*{10pt}

\vspace*{3pt}

\vspace*{1pt}\textlineskip    

\section{Introduction} \label{introduction} 

\noindent
Quantum walks (QWs) are quantum versions of classical random walks, and they have attracted much attention in various fields. In particular, QWs are expected to provide mathematical models for quantum algorithms, which can be implemented in quantum computers \cite{basic1,basic2,basic3,basic4}. Thus, various types of QWs have been analyzed since the early 2000s. In this work, we consider QWs starting from the origin on the integer lattice, where the walker moves to the left and to the right. The time evolution of QWs is defined with the coin operator and the shift operator.  Also, we assume that the coin operator is position-independent, i.e., the homogeneous coin operator. Usually, the time evolution of  QWs is linear and unitary since, at each position, the coin operator is defined as a unitary matrix called the coin matrix. To distinguish these common QWs from the models that we will introduce later, we call them linear quantum walks (LQWs). Despite the simplicity of the definition, LQWs show a remarkable spreading property that classical random walks do not possess. The standard deviation of the walker’s position grows linearly in time and quadratically faster than classical random walks. Konno \cite{Konno2002,Konno2005} proved this phenomenon rigorously by deriving a convergence in the distribution of a rescaled walker’s position (weak limit theorem) of the 2-state LQW for an arbitrary coin operator. Likewise, our focus is to derive a weak limit theorem, which can help us visualize the outlines of the probability distributions of a walker’s position.\\
\indent This article aims to introduce a formalism of decompose-type quantum walks (DQWs), which are the extensions of LQWs. DQWs have the non-linear isometric coin operator, which consists of 2 different matrices, where at each position, one is assigned to a real component, and the other is assigned to an imaginary component of the quantum state. Our primary motivation for introducing DQWs is to investigate the behavior of non-linear QWs, which is realized by increasing the degrees of freedom. By regarding the real and imaginary parts as different elements of the quantum state, without loss of information about quantum states, we can convert 2-state DQWs to 4-state LQWs. The initial state of these 4-state LQWs is restricted in $\mathbb{R}^4$ instead of $\mathbb{C}^4$. Here, $\mathbb{R}$ is the set of real numbers, and $\mathbb{C}$ is the set of complex numbers. The probability distributions of a walker's position in the DQW and their corresponding 4-state LQW are identical. In other words, we can achieve a class of 4-state LQWs by DQWs with only 2 states. Quaternionic quantum walks, which are the extensions of 2-state LQWs, can also be regarded as 4-state LQWs \cite{quaternionic1,quaternionic2}, and other kinds of non-linear quantum walks have also been proposed and analyzed \cite{nonlinear1,nonlinear2,nonlinear3,nonlinear4,nonlinear5,nonlinear6,nonlinear7}.  Furthermore, many types of LQWs with 3 or more states have been investigated \cite{multi1,multi3,multi4,multi5,multi6,multi7}. In the study of multi-state LQWs, the Grover walk often plays an important role. The name comes from Grover's search algorithm \cite{search1}, and it has the Grover matrix as a coin matrix. Grover walks are expected to speed up the search algorithms on general graphs \cite{search2}. Additionally, localization is a striking property of Grover walks, and it means that the probability measure takes a positive value at a certain position in the limit of infinite time. While only trivial models exhibit localization for 2-state LQWs, some non-trivial models exhibit localization for multi-state LQWs, including the Grover walks. The probability distributions of these models rescaled by time converges weakly to a random variable, which has the Dirac delta function and a continuous function with finite support. The Dirac delta function contributes to localization, and the continuous function corresponds to the spreading phenomenon. Interestingly, localization and the spreading phenomenon can occur simultaneously in multi-state LQWs. The 3-state and 4-state Grover walks were studied thoroughly in \cite{inui2005one,inui2005localization}, and Machida \cite{machida2014limit} derived the weak limit theorem for a one-parameter family of the 3-state Grover walk. However, the result for a one-parameter family of the 4-state Grover walk is not obtained. In this paper, we reveal that localization and the spreading phenomenon can coincide in DQWs, and there is a DQW corresponding to the 4-state Grover walk (note that the shift operator is slightly different from \cite{inui2005localization}). Thus, we derive the weak limit theorem of a one-parameter family of the 4-state Grover walk generated by DQWs.\\
\indent The rest of this paper is organized as follows. In Section \ref{LQWs}, we introduce 2-state LQWs and the weak limit theorem. In Section \ref{DQWs}, we show the formalism of DQWs and give a method to convert them to corresponding 4-state LQWs. We also introduce a generalized 4-state Grover walk generated by DQWs. Section \ref{main} is devoted to the analysis of 4-state LQWs by the Fourier transform. We utilize the method introduced by Grimmett et al. \cite{GJS} to derive the weak limit theorem. Finally, we compare the probability distributions and the limit density functions by computer simulations to demonstrate the validity of our results.

\section{LQWs\label{LQWs}}
We introduce 2-state LQWs on the integer lattice. Firstly, we consider the Hilbert space defined as
$\mathcal{H}_{\mathbb{C}^{2}}=\ell^{2}\left(\mathbb{Z} ; \mathbb{C}^{2}\right)=\left\{\Psi: \mathbb{Z} \rightarrow \mathbb{C}^{2} | \sum_{x \in \mathbb{Z}}\|\Psi(x)\|^{2}_{\mathbb{C}^2}<\infty\right\},$ where $\mathbb{Z}$ is the set of integers. The quantum state $\Psi$ is an element of $\mathcal{H}_{\mathbb{C}^{2}}$. Since we concentrate on the case that the walker departs from the origin, the initial state $\Psi_{0} \in \mathcal{H}_{\mathbb{C}^{2}}$ is given with a non-zero value at the origin:

\[
\Psi(x)=\left(\begin{array}{l}
\Psi_{1}(x) \\
\Psi_{2}(x)
\end{array}\right)
,\quad
\Psi_0(x)
=\begin{pmatrix}
\phi_1 \\ \phi_2
\end{pmatrix}
\delta_0(x)
,\]
where $\delta_{0}(x) (x\in\mathbb{Z} )$  is a function which outputs 1 at the origin and 0 at the other positions. It is generally formulated as follows for the sake of the rest of this article:

 \begin{align}\label{Delta function}
 \delta_{x_0}(x)=\left\{\begin{array}{ll}
1 , & x=x_0,\\
0, & x \neq x_0.
\end{array}\right.
 \end{align}
 
The time evolution operator of the quantum state is formulated as  $U=S C$, where $S$ denotes the shift operator and $C$ denotes the coin operator. Both of them are operators on $\mathcal{H}_{\mathbb{C}^{2}} $, thus $U$  is an operator on $\mathcal{H}_{\mathbb{C}^{2}} $ as well. The coin operator of the LQW is defined by the coin matrix acting on the quantum state at each position. In this paper, we assume that the coin operator is homogeneous; that is, its coin matrix is given as a position-independent $2 \times 2$ unitary matrix $M$:
\[M=\left(\begin{array}{ll}
a & b \\
c & d
\end{array}\right),
\quad (C \Psi)(x)=M \Psi(x).\]
Additionally, the shift operator $S$ is defined by
\[
\ 
(S \Psi)(x)=\left(\begin{array}{l}
\Psi_{1}(x+1) \\
\Psi_{2}(x-1)
\end{array}\right)
. \] It shifts the first element of $\Psi(x)$ to the left and the second element to the right. The time evolution operator can also be formulated by decomposing the coin matrix $M$ into $P$ and $Q$, where the first and second rows of $M$ are pulled out, respectively. Then, $U$ can be rewritten only by $P$ and $Q$:
\[P=\left(\begin{array}{ll}
a & b \\
0 & 0
\end{array}\right),
\quad
Q=\left(\begin{array}{ll}
0 & 0 \\
c & d
\end{array}\right),
\quad
(U \Psi)(x)=P \Psi(x+1)+Q \Psi(x-1).\]
Finally, we define the probability measures of the LQW at time $n$ by the $L^{2}$ norm as below:
\[\mu_{n}(x)=\left\|\left(U^{n} \Psi_{0}\right)(x)\right\|^{2}=\left|\left(U^{n} \Psi_{0}\right)_{1}(x)\right|^{2}+\left|\left(U^{n} \Psi_{0}\right)_{2}(x)\right|^{2}.\]Here we also define localization by `` there exist $\Psi_0\in\mathcal{H}_{\mathbb{C}^2}$ and $x \in \mathbb{Z}$ such that $ \limsup _{n \rightarrow \infty} $ $\mu_{n}(x)>0$ ". We show three examples of the probability distributions at time $n=100$  with the initial state $\Psi_0(x) = {}^T(\begin{array}{l}\frac{1}{\sqrt{2}} , \frac{i}{\sqrt{2}}\end{array})\delta_0(x)$ in Fig. \ref{2-state LQWs Figure}. With respect to Figure \ref{2-state LQWs Figure}-(a), when time $n$ is even, the probability measure is 1 at the origin, and when $n$ is odd, the probability measures are 0.5 at position $\pm 1$, respectively. This  LQW satisfies the condition of localization,  $\limsup _{n \rightarrow \infty} \mu_{n}(0)>$
0. Thus, we call this model the trivial LQW model that exhibits localization. On the other hand, from Figs. \ref{2-state LQWs Figure}-(b) and \ref{2-state LQWs Figure}-(c), we can see that the quantum walkers are spreading instead of staying at the origin. Most of the distributions of 2-state LQWs show an inverted bell-shaped form similar to (c).

\begin{figure}[htbp]
\captionsetup{justification=centering}
\begin{subfigure}{0.326\textwidth}
\includegraphics[width=1\linewidth, height=3.5cm]{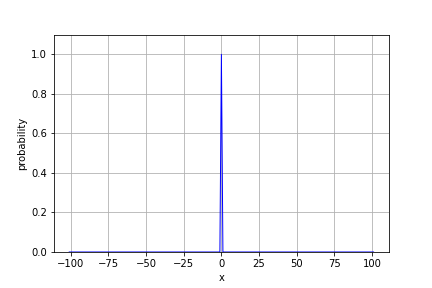} 
\subcaption{\small{$M=\left(\begin{array}{ll}
0 & 1 \\
1 & 0
\end{array}\right)$}}
\end{subfigure}
\begin{subfigure}{0.326\textwidth}
\includegraphics[width=1\linewidth, height=3.5cm]{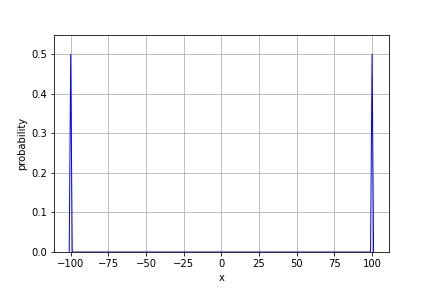}
\subcaption{\small{$M=\left(\begin{array}{ll}
1 & 0 \\
0 & 1
\end{array}\right)$}}

\end{subfigure}
\begin{subfigure}{0.326\textwidth}
\includegraphics[width=1\linewidth, height=3.5cm]{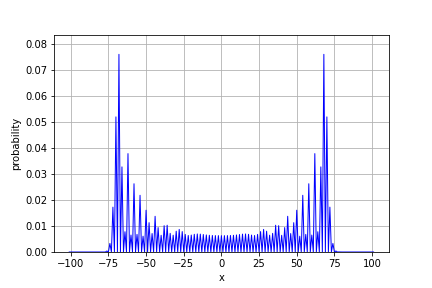} 
\subcaption{\small{$M=\frac{1}{\sqrt{2}}\left(\begin{array}{ll}
1 & 1 \\
1 & -1
\end{array}\right)$}}
\end{subfigure}
\vspace*{13pt}
\addtocounter{figure}{-1} 
\fcaption{Three examples of the probability distributions of 2-state LQWs at time $n=100$  with the initial state $\Psi_0(x) = {}^T(\begin{array}{l}\frac{1}{\sqrt{2}} , \frac{i}{\sqrt{2}}\end{array})\delta_0(x)$. }
\medskip
\label{2-state LQWs Figure}
\end{figure}

The weak limit theorem for 2-state LQWs was given in \cite{Konno2002,Konno2005}
by a path counting method.
Let $X_{n}$ be a random variable defined by \(\mathbb{P}\left(X_{n}=x\right)=\mu_{n}(x)\). Since LQWs spread quadratically faster than classical random walks, we rescale the probability distribution by $n$ instead of $\sqrt{n}$. The weak limit theorem provides a limit distribution of the spatially rescaled random value $X_n/n$. 

\vspace*{12pt}
\begin{theorem}\label{konno theorem}
\emph{(\text {Konno} \cite{Konno2002,Konno2005})}

\[ \lim _{n \rightarrow \infty} \mathbb{P}\left(\frac{X_{n}}{n} \leq x\right)=\int_{-\infty}^{x}\left\{1-C\left(a, b ; \phi_{1}, \phi_{2}\right) y\right\} f_{K}(y ;|a|)dy,\]
where 
\begin{align*}
& C\left(a, b ; \phi_{1}, \phi_{2}\right)=\left|\phi_{1}\right|^{2}-\left|\phi_{2}\right|^{2}-\frac{a \phi_{1} \overline{b \phi_{2}}+\overline{a \phi_{1}} b \phi_{2}}{|a|^{2}},
 \\
&f_{K}(x ; r)=\frac{\sqrt{1-r^{2}}}{\pi\left(1-x^{2}\right) \sqrt{r^{2}-x^{2}}} I_{(-r, r)}(x) \ (0<r<1),
\\
&I_{(-r, r)}(x)=\left\{\begin{array}{ll}
1, & (x \in(-r, r)) \\
0, & (x \notin(-r, r))
\end{array}\right..
\end{align*}

\end{theorem}
Here $f_{K}(x ; r)$ is a limit density function of 2-state LQWs spatially rescaled by time $n$, which contributes to an inverse bell-shaped form of the probability distribution (Fig. \ref{2-state LQWs Figure}-(c)). It has a compact support, and its domain is $(-|a|,|a|)$, where $a$ is the $(1,1)$  entry of the coin matrix $M$.  

\section{DQWs\label{DQWs}}
In this section, we focus on introducing the definition of DQWs. The difference between LQWs and DQWs is in their coin operators. Subsequently, we convert DQWs to corresponding 4-state LQWs and define a one-parameter family of the 4-state Grover walk generated from DQWs.
\subsection{Definition of DQWs}
We consider the same quantum state $\Psi \in \mathcal{H}_{\mathbb{C}^{2}}$. However, for the DQWs case, we decompose the quantum state into real and imaginary components: 
\[\Psi(x)=\left(\begin{array}{l}
\Psi_{1}(x) \\
\Psi_{2}(x)
\end{array}\right)=(\Re \Psi)(x)+(\Im \Psi)(x) i.\]
Here $\Re$ and $\Im$ are operators on $\mathcal{H}_{\mathbb{C}^{2}}$, and they pull out the real and imaginary components of the quantum state:
\[(\Re \Psi)(x)=\left(\begin{array}{c}
\Re\left(\Psi_{1}(x)\right) \\
\Re\left(\Psi_{2}(x)\right)
\end{array}\right),
\quad
(\Im \Psi)(x)=\left(\begin{array}{c}
\Im\left(\Psi_{1}(x)\right) \\
\Im\left(\Psi_{2}(x)\right)
\end{array}\right) \in \mathbb{R}^{2},\]where $\Re(z)$ and $\Im(z)$ are the real and imaginary parts of a complex number $z$, respectively. Since we treat the walk starting from the origin, the initial state can also be decomposed as below:
\begin{align*}
(\Re\Psi_0)(x)
=\begin{pmatrix}
\Re{(\phi_1)} \\ \Re{(\phi_2)}
\end{pmatrix}\delta_{0}(x),
\quad
(\Im\Psi_0)(x)
=
\begin{pmatrix}
\Im{(\phi_1)} \\ \Im{(\phi_2)}
\end{pmatrix}
\delta_{0}(x).
\end{align*}Let $U_D$ denote the time evolution operator given as $U_D=SC_D$. Let $C_{D} $ denote the coin operator of DQWs, and it is formulated with two $2 \times 2$ matrices $M_{R}$ and $M_{I}$  corresponding to $(\Re\Psi)(x)$ and $\left(\Im \Psi\right)(x)$, respectively. We assume $a_{R}, b_{R}, c_{R}, d_{R}, a_{I}, b_{I}, c_{I}, d_{I} \in \mathbb{C},$ then
\[\left(C_{D} \Psi\right)(x)=M_{R}(\Re \Psi)(x)+M_{I}(\Im \Psi)(x) i,\]where
\[\ M_{R}=\left(\begin{array}{cc}
a_{R} & b_{R} \\
c_{R} & d_{R}
\end{array}\right),\quad M_{I}=\left(\begin{array}{ll}
a_{I} & b_{I} \\
c_{I} & d_{I}
\end{array}\right).\]
The time evolution operator  $U_{D}$  can also be formulated by decomposing $M_{R}$ and $M_{I}$ into $P_{R}, Q_{R}$ and $P_{I}, Q_{I}$, respectively:
\[P_{R}=\left(\begin{array}{ll}
a_{R} & b_{R} \\
0 & 0
\end{array}\right),\quad  
Q_{R}=\left(\begin{array}{ll}
0 & 0 \\
c_{R} & d_{R}
\end{array}\right),\quad 
P_{I}=\left(\begin{array}{ll}
a_{I} & b_{I} \\
0 & 0
\end{array}\right),\quad  Q_{I}=\left(\begin{array}{ll}
0 & 0 \\
c_{I} & d_{I}
\end{array}\right),\]
\[\left(U_{D} \Psi\right)(x)=P_{R}(\Re \Psi)(x+1)+Q_{R}(\Re \Psi)(x-1)+\left\{P_{I}(\Im \Psi)(x+1)+Q_{I}(\Im \Psi)(x-1)\right\} i.\]Note that the coin operator $U_D$ is no longer unitary for general. However, to preserve a sum of the probability measures of the DQW, we consider the necessary and sufficient condition for $C_D$ to be an isometry. 
\vspace*{12pt}
\begin{lemma}\label{coin pair}
The operator $C_D$ is an isometry $\Leftrightarrow$ $M_{R}, M_{I}$ are unitary matrices and $M_{R}^{*} M_{I}$ is a real matrix. Here, ``$\Leftrightarrow$" means``if and only if".
\begin{proof}Put $\psi \in \mathbb{C}^2$, and we decompose $\psi$ into the real and imaginary components, i.e., $\psi_R$ and $\psi_I$:
\[\psi=\psi_{R}+\psi_{I} i, \quad \psi_{R}=\left(\begin{array}{c}
\alpha_{R} \\
\beta_{R}
\end{array}\right),\quad 
\psi_{I}=\left(\begin{array}{c}
\alpha_{I} \\
\beta_{I}
\end{array}\right) \in \mathbb{R}^{2}.\]The coin operator $C_D$ is an isometry, i.e., $\left\|C_{D} \Psi\right\|_{\mathcal{H}_{\mathbb{C} 2}}^{2}=\|\Psi\|_{\mathcal{H}_{\mathbb{C} 2}}^{2}$ for any $\Psi \in \mathcal{H}_{\mathbb{C}^{2}}$ $(\Longleftrightarrow\left\|\left(C_{D} \Psi\right)(x)\right\|^{2} =\|\Psi(x)\|^{2}$
for any $ x \in \mathbb{Z}$ and $\Psi \in \mathcal{H}_{\mathbb{C}^{2}}$). This is equivalent to the condition below:
\begin{align}
\label{mainEquation}
\left\|M_{R} \psi_{R}+M_{I} \psi_{I} i\right\|^{2}=\|\psi\|^{2}\ \text{for any}\ \psi \in \mathbb{C}^{2},
\end{align}
where
\begin{align}
\label{one in lemma}
||M_R\psi_R+M_I\psi_I i ||^2
&=\bra{\psi_R}M_R^*M_R\ket{\psi_R}+\bra{\psi_I}M_I^*M_I\ket{\psi_I}-2\Im(\bra{\psi_R}M_R^*M_I\ket{\psi_I}),
\\
\label{two in lemma}
 ||\psi||^2&=||\psi_R||^2+||\psi_I||^2.
\end{align}
 The sufficiency $(\Leftarrow)$ of this lemma is obvious, thus we prove the necessity $(\Rightarrow)$ henceforward. We assume that (\ref{mainEquation}) holds. When $\psi_{I}=^{T}(0,0),$ then $\left\langle\psi_{R}\left|M_{R}^{*} M_{R}\right| \psi_{R}\right\rangle=\left\langle\psi_{R} | \psi_{R}\right\rangle $. When $\psi_{R}=^{T}(0,0)$, then $\left\langle\psi_{I}\left|M_{I}^{*} M_{I}\right| \psi_{I}\right\rangle=\left\langle\psi_{I} | \psi_{I}\right\rangle .$ Thus, $ M_{R}^{*} M_{R}$ and $M_{I}^{*} M_{I}$ should be identity matrices, and $\Im\left(\left\langle\psi_{R}\left|M_{R}^{*} M_{I}\right| \psi_{I}\right\rangle\right)$ should be $0$  for any $\psi$. By calculating $\left\langle\psi_{R}\left|M_{R}^{*} M_{I}\right| \psi_{I}\right\rangle$ specifically, we get
\[\begin{aligned}
\Im\left(\left\langle\psi_{R}\left|M_{R}^{*} M_{I}\right| \psi_{I}\right\rangle\right)=0 
& \Leftrightarrow\left\langle\psi_{R}\left|M_{R}^{*} M_{I}\right| \psi_{I}\right\rangle \in \mathbb{R} 
\\
&\Leftrightarrow\left\langle\psi_{R}\left|M_{R}^{*} M_{I}\right| \psi_{I}\right\rangle-\overline{\left\langle\psi_{R}\left|M_{R}^{*} M_{I}\right| \psi_{I}\right\rangle}=0
\\
&\Leftrightarrow 2i( \alpha _{I} \alpha _{R}\left( \Im \left( a_{I}\overline{a_{R}}\right) +\Im \left( c_{I}\overline{c_{R}}\right)\right) +\alpha _{I} \beta _{R}\left( \Im \left( a_{I}\overline{b_{R}}\right) +\Im \left( c_{I}\overline{d_{R}}\right)\right)
\\
&+\alpha _{R} \beta _{I}\left( \Im \left( b_{I}\overline{a_{R}}\right) +\Im \left( d_{I}\overline{c_{R}}\right)\right) +\beta _{I} \beta _{R}\left( \Im \left( b_{I}\overline{b_{R}}\right) +\Im \left( d_{I}\overline{d_{R}}\right)\right))=0.
\end{aligned}\]Hence,  
\[\Im\left(\overline{a_{R}} a_{I}+\overline{c_{R}} c_{I}\right)=\Im\left(\overline{a_{R}} b_{I}+\overline{c_{R}} d_{I}\right)=\Im\left(\overline{b_{R}} \alpha_{I}+\overline{d_{R} c_{I}}\right)=\Im\left(\overline{b_{R}} b_{I}+\overline{d_{R}} d_{I}\right)=0\]holds, i.e, each entry of $M_{R}^{*} M_{I}$ is a real number. Hence, both of $M_R$ and $M_I$ are unitary matrices, and $M_{R}^{*} M_{I}$ is a real matrix. Therefore, we prove the necessity of this lemma.
\end{proof}
\end{lemma}

From Lemma \ref{coin pair}, we assume that matrices $M_{R}$ and $ M_{I}$ are unitary matrices, and $M_{R}^{*} M_{I}$ is an orthogonal matrix. In this paper, we consider a pair of matrices $(M_R,M_I)$ satisfying

\[M_{R}^{*} M_{I}=\left(\begin{array}{cc}e & f \\ f & -e\end{array}\right),\quad \left(e, f \in \mathbb{R},\  e^{2}+f^{2}=1\right).\]
Hence, $M_{R}$ and $M_{I}$ can be expressed with $\Delta \in[-\pi, \pi), \alpha, \beta \in \mathbb{C}$ as

\begin{align}
\label {mr mi}
M_{R}=e^{i \Delta}\left(\begin{array}{cc}
\alpha & \beta \\
-\bar{\beta} & \bar{\alpha}
\end{array}\right),\quad  M_{I}=e^{i \Delta}\left(\begin{array}{cc}
e \alpha+f \beta & f \alpha-e \beta \\
-e \bar{\beta}+f \bar{\alpha} & -f \bar{\beta}-e \bar{\alpha}
\end{array}\right),
\end{align}
where $|\alpha|^{2}+|\beta|^{2}=1 .$ Finally, the probability measure at time $n$ is obtained as

\[\mu_{n}(x)=\left\|\left(U^{n} \Psi_{0}\right)(x)\right\|^{2}=\left\|\left(\Re U^{n} \Psi_{0}\right)(x)\right\|^{2}+\left\|\left(\Im U^{n} \Psi_{0}\right)(x)\right\|^{2}.\]When both $M_{R}$ and $M_{I}$ are orthogonal matrices, Lemma \ref{coin pair} implies that $U_{D}$ is an isometry. Then, DQWs satisfy
\begin{align*}
&P_{R}(\Re \Psi)(x+1)+Q_{R}(\Re \Psi)(x-1) \in \mathbb{R}^{2}, \\
&P_{I}(\Im \Psi)(x+1)+Q_{I}(\Im \Psi)(x-1) \in \mathbb{R}^{2}.
\end{align*}
Thus, we can consider the time evolution for $\Re \Psi$ and $\Im \Psi$ independently, which means
\begin{align*}
\left(\Re U_{D} \Psi\right)(x)&=P_{R}(\Re \Psi)(x+1)+Q_{R}(\Re \Psi)(x-1),
\\
\left(\Im U_{D} \Psi\right)(x)&=P_{I}(\Im \Psi)(x+1)+Q_{I}(\Im \Psi)(x-1).\end{align*}Therefore, the probability measure $\mu_{n}(x)$ becomes a sum of the probability measures of two LQWs whose coin matrices are $M_{R}$ and $M_{I}$ with the initial states $\Re \Psi_{0}$ and\ $\Im \Psi_{0}$, respectively. Moreover, the ratio of the probability measures of these LQWs is determined by  the amounts of $||\Re \Psi_{0}||^2$ and $||\Im \Psi_{0}||^2$, respectively. We show three examples of $\mu_{100}(x)  $ of DQWs with the pairs of real matrices  in Fig. \ref{DQWs Figure}. From these examples, we can see that the probability distributions seem to be the overlaps of the distributions in Fig. \ref{2-state LQWs Figure}. However, note that since  $\Re \Psi_{0} ,\Im \Psi_{0}\in \mathbb{R}^2$, there are no initial states which make the probability distributions symmetric for Figs. \ref{DQWs Figure}-(b) and \ref{DQWs Figure}-(c) (see \cite{Konno2002}). Instead, we set the initial states for Figs. \ref{DQWs Figure}-(b) and \ref{DQWs Figure}-(c) which make their limit density functions (Theorem \ref{konno theorem}) symmetric.

\begin{figure}[h]

\begin{subfigure}{0.325\textwidth}
\centering
\includegraphics[width=1\linewidth, height=3.5cm]{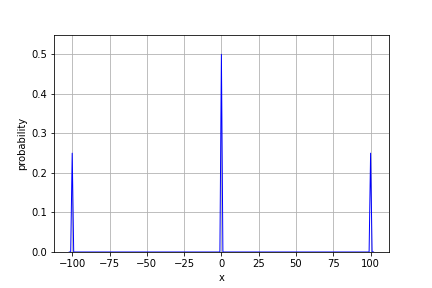} 
\caption{}
\end{subfigure}
\begin{subfigure}{0.325\textwidth}
\centering
\includegraphics[width=1\linewidth, height=3.5cm]{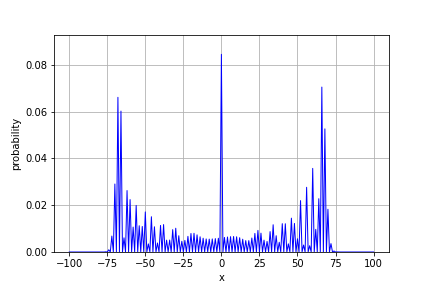}
\caption{}

\end{subfigure}
\begin{subfigure}{0.325\textwidth}
\centering
\includegraphics[width=1\linewidth, height=3.5cm]{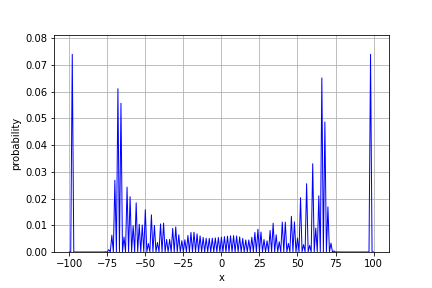} 
\caption{}
\end{subfigure}
\addtocounter{figure}{-1} 
\fcaption{Three examples of the probability distributions of DQWs at time $n=100$ with pairs of real matrices $(M_R,M_I)$ and initial states $\Psi_0(0)$.}
\medskip
\label{DQWs Figure}

\small{\begin{align*}
( a) & \ M_{R} =\left(\begin{array}{ l l }
1 & 0\\
0 & 1
\end{array}\right),
\quad 
M_{I} =\left(\begin{array}{ l l }
0 & 1\\
1 & 0
\end{array}\right) ,\quad
\Psi _{0} (0)=\frac{1}{2}\left(\begin{array}{ c }
1+i\\
1+i
\end{array}\right),
\\
( b) & \ M_{R} =\frac{1}{\sqrt{2}}\left(\begin{array}{ l l }
1 & \ \ 1\\
1 & -1
\end{array}\right) ,
\quad 
M_{I} =\left(\begin{array}{ l l }
0 & 1\\
1 & 0
\end{array}\right),
\quad
\Psi _{0} (0)=\frac{1}{25\sqrt{2}}\left(\begin{array}{ c }
24\sqrt{2-\sqrt{2}} +7\sqrt{2} i\\
24\sqrt{2+\sqrt{2}} +7\sqrt{2} i
\end{array}\right),
\\
( c) & \ M_{R} =\frac{1}{\sqrt{2}}\left(\begin{array}{ l l }
1 & \ \ 1\\
1 & -1
\end{array}\right) ,
\quad
M_{I} =\left(\begin{array}{ l l }
1 & 0\\
0 & 1
\end{array}\right),
\quad
\Psi _{0} (0)=\frac{1}{13\sqrt{2}}\left(\begin{array}{ c }
12\sqrt{2-\sqrt{2}} +5\sqrt{2} i\\
12\sqrt{2+\sqrt{2}} +5\sqrt{2} i
\end{array}\right).
\end{align*}}
\end{figure}

\subsection{ DQWs to 4-state LQWs}
When $M_{R}$ or $M_{I}$ is not a real matrix, the time evolution cannot be considered individually for $\Re \Psi$ and $ \Im \Psi$. Therefore, we analyze the corresponding 4-state LQW, which has the identical probability distribution. Firstly, we define an operator $\iota: \mathcal{H}_{\mathbb{C}^{2}} \rightarrow \mathcal{H}_{\mathbb{R}^{4}}$ to get the 4-state quantum state corresponding to $\Psi$. Also, the corresponding coin matrix $\tilde{M}$ is given by $M_{R}$ and $M_{I}$ as a $4 \times 4$ matrix:
\[(\iota \Psi)(x)=\left(\begin{array}{c}
\Re\left(\Psi_{1}(x)\right) \\
\Re\left(\Psi_{2}(x)\right) \\
\Im\left(\Psi_{1}(x)\right) \\
\Im\left(\Psi_{2}(x)\right)
\end{array}\right),
\quad
\tilde{M}=\left(\begin{array}{l|r}
{\rm \Re} M_R &-{\rm \Im} M_I \\ \hline
{\rm \Im} M_R &{\rm \Re} M_I 
\end{array}\right),\]
where $\Re$ and $\Im$ pull out the real and imaginary components of each matrix, respectively, that is, for all complex numbers $j_1,\ j_2,\ j_3\ \text{and}\ j_4$, $\Re$ and $\Im$ satisfy

\[\Re\left(\begin{array}{cc}
j_{1} & j_{2} \\
j_{3} & j_{4}
\end{array}\right)=\left(\begin{array}{cc}
\Re\left(j_{1}\right) & \Re\left(j_{2}\right) \\
\Re\left(j_{3}\right) & \Re\left(j_{4}\right)
\end{array}\right),
\quad
\Im\left(\begin{array}{cc}
j_{1} & j_{2} \\
j_{3} & j_{4}
\end{array}\right)=\left(\begin{array}{cc}
\Im\left(j_{1}\right) & \Im\left(j_{2}\right) \\
\Im\left(j_{3}\right) & \Im\left(j_{4}\right)
\end{array}\right).\]
  Note that $\iota$ is a unitary map  since $\iota\iota^*=I_{\mathcal{H}_{\mathbb{R}^4}}$ and $\iota^*\iota=I_{\mathcal{H}_{\mathbb{C}^2 }}$ hold, where the adjoint of $\iota$ is given as $\left(\iota^{*} \psi\right)(x)=\begin{pmatrix}
\psi _{1} (x)+i\psi _{3} (x) & \psi _{2} (x)+i\psi _{4} (x)
\end{pmatrix}^{\top}$ for any $\psi \in \mathcal{H}_{\mathbb{R}^{4}}$ with $\psi(x)=\begin{pmatrix}
\psi _{1}( x) & \psi _{2}( x) & \psi _{3}( x) & \psi _{4}( x)
\end{pmatrix}^{\top}$. Under the assumptions of (\ref{mr mi}), by putting $\alpha=|\alpha|e^{i\theta}$ and $\beta=|\beta|e^{i\phi}$, $\tilde{M}$ can be expressed specifically as
\[\scriptsize\left(\begin{array}{cccc}
|\alpha| \cos (\theta+\Delta) 
& |\beta| \cos (\phi+\Delta) 
& -e|\alpha| \sin (\theta+\Delta)-f|\beta| \sin (\phi+\Delta) 
& e|\beta| \sin (\phi+\Delta)-f|\alpha| \sin (\theta+\Delta) 
\\
-|\beta| \cos (\phi-\Delta) 
& |\alpha| \cos (\theta-\Delta) 
& -e|\beta| \sin (\phi-\Delta)+f|\alpha| \sin (\theta-\Delta)
& -e|\alpha| \sin (\theta-\Delta)-f|\beta| \sin (\phi-\Delta) 
\\
|\alpha| \sin (\theta+\Delta) 
& |\beta| \sin (\phi+\Delta)
& e|\alpha| \cos (\theta+\Delta)+f|\beta| \cos (\phi+\Delta) 
& -e|\beta| \cos (\phi+\Delta)+f|\alpha| \cos (\theta+\Delta) 
\\
|\beta| \sin (\phi-\Delta) 
& -|\alpha| \sin (\theta-\Delta) 
& -e|\beta| \cos (\phi-\Delta)+f|\alpha| \cos (\theta-\Delta)
& -f|\beta| \cos (\phi-\Delta)-e|\alpha| \cos (\theta-\Delta)
\end{array}\right).\]The time evolution operator is defined by $\tilde{U}=\tilde{S}\tilde{C}$, where the coin operators $\tilde{C}$ and the shift operator $\tilde{S}$ are determined by 
\[(\tilde{C} \iota \Psi)(x)=\tilde{M}(\iota \Psi)(x),
\quad (\tilde{S} \iota \Psi)(x)=\left(\begin{array}{c}
\Re\left(\Psi_{1}(x+1)\right) \\
\Re\left(\Psi_{2}(x-1)\right) \\
\Im\left(\Psi_{1}(x+1)\right) \\
\Im\left(\Psi_{2}(x-1)\right)
\end{array}\right).\]By decomposing $\tilde{M}$ into $\tilde{P}$ and $\tilde{Q}$ as
\[\tilde{P}=\left(\begin{array}{cccc}
1 & 0 & 0 & 0 \\
0 & 0 & 0 & 0 \\
0 & 0 & 1 & 0 \\
0 & 0 & 0 & 0
\end{array}\right) \tilde{M},
\quad \tilde{Q}=\left(\begin{array}{cccc}
0 & 0 & 0 & 0 \\
0 & 1 & 0 & 0 \\
0 & 0 & 0 & 0 \\
0 & 0 & 0 & 1
\end{array}\right) \tilde{M},\]the time evolution operator $\tilde{U}$ can also be expressed by $\tilde{P}$ and $\tilde{Q}$ as below:
\[
(\tilde{U}\iota\Psi)(x)=\tilde{P}\iota\Psi(x+1)
+
\tilde{Q}\iota\Psi(x-1).
\]
Since $
\iota S \iota^{*}=\tilde{S}
$ and $
\iota C_{D} \iota^{*}=\tilde{C}
$, and $\iota$ is a unitary map, then $
\tilde{U}=\iota U_{D} \iota^{*}
$ holds. This implies $
\tilde{U}^{n} \iota \Psi_{0}=\iota U_{D}^n\Psi_0
$ for any $n$, and also
\[\mu_n(x)=\|(U_{D}^n\Psi_0)(x)\|_{\mathbb{C}^2}^2=\|(\tilde{U}^n\iota\Psi_0)(x)\|_{\mathbb{R}^4}^2.\]
Here we also define the quantum state at time $n$ in the Fourier space $\hat{\Psi}_n \in \ell^2([-\pi,\pi)\rightarrow\mathbb{C}^4:\frac{dk}{2\pi})$. It is given as the Fourier transform of $\tilde{U}^n\iota\Psi$, which is the quantum state at time $n$:
\[\hat{\Psi}_n(k)=\displaystyle\sum_{x\in \mathbb{Z}}e^{-ikx}(\tilde{U}^n\iota\Psi)(x).\]
From the inverse Fourier transformation, we have
\[(\tilde{U}^n\iota\Psi)(x)=\displaystyle\int_{-\pi}^{\pi}e^{ikx}\hat{\Psi}_n(k)\frac{dk}{2\pi}.\]
 By defining a $4\times 4$ matrix $\hat{U}(k)$ as
\[\hat{U}(k)=\left(\begin{array}{cccc}
e^{i k} & 0 & 0 & 0 \\
0 & e^{-i k} & 0 & 0 \\
0 & 0 & e^{i k} & 0 \\
0 & 0 & 0 & e^{-i k}
\end{array}\right) \tilde{M},
\]
the time evolution of $\hat{\Psi}_n(k)$ can be obtained in a simple way:
\[\hat{\Psi}_n(k)=\hat{U}(k)\hat{\Psi}_{n-1}(k).\]
As we mentioned in Section \ref{introduction}, the Grover walks are particularly important. The coin matrix of the 4-state Grover walk is given by the following 4 $\times$ 4 Grover matrix:\[
\frac{1}{2}\left(\begin{array}{cccc}
-1 & 1 & 1 &1 \\
1 &-1 & 1 & 1 \\
1& 1 & -1 & 1 \\
1 &1 &1 &-1
\end{array}\right)
.\]One of the most remarkable properties of Grover walks is localization, and it is caused by eigenvalues 1 and -1 of the time evolution operator in the Fourier space (see \cite{suzuki2016asymptotic}). Thus, to consider the generalized Grover walk, we obtain the necessary and sufficient conditions for $\hat{U}(k)$  to have eigenvalues 1 and -1. Under these conditions, we prove the weak limit theorem of 4-state LQWs corresponding to DQWs whose $\tilde{M}$ includes the Grover matrix. The next lemma is obtained by directly analysing the characteristic polynomial of $\hat{U}(k)$.
\begin{lemma}\label{eigenvalue lemma}
 $1$ and $-1$ are eigenvalues of $\hat{U}(k)$ if and only if one of the following conditions holds:

\begin{enumerate}
\item $ \alpha=0\  {\rm and} \ b\cos\phi \cos\Delta=0,$
\item $ \beta=0,\ a=0\ and \ \sin\theta \sin\Delta=0,$
\item $ \alpha,\beta,b\neq0, \ \cos(\theta-\phi)=-\frac{a|\alpha|}{b|\beta|}\  {\rm and} \ \sin\theta=0,$
\item $\alpha,\beta,b\neq0, \ \cos(\theta-\phi)=-\frac{a|\alpha|}{b|\beta|}, \ \cos\Delta\neq 0\   {\rm and} \ \tan\Delta=\frac{|\alpha|a\cos\theta+b|\beta|\cos\phi}{|a|\sin\theta}.$
\end{enumerate}

\begin{proof}
The characteristic polynomial of $\hat{U}(k)$ is described as
\[x^{4}+A x^{3}+2 i|\alpha| \sin (2 k)(e|\alpha|+f|\beta| \cos (\theta-\phi)) x^{2}-\bar{A} x-1,\]where 
\[\begin{aligned}
A=&-2 \cos (k)\{|\alpha| \cos (\Delta) \cos (\theta)-e \sin (\Delta) \sin (\theta))-f|\beta| \sin (\Delta) \sin (\phi)\} 
\\
&-2 i \sin (k)\{|\alpha|(e \cos (\Delta) \cos (\theta)-\sin (\Delta) \sin (\theta))+f|\beta| \cos (\Delta) \cos (\phi)\}.
\end{aligned}\] Then, 1 and -1 are eigenvalues of  $\widehat{U}(k)$ if and only if the following condition holds:
\begin{align*}
  &A-\bar{A}+2 i|\alpha| \sin (2 k)(e|\alpha|+f|\beta| \cos (\theta-\phi))=0 \ (\text{for any}\ k\in [-\pi,\pi))
\\
&\Leftrightarrow|\alpha|(e|\alpha|+f|\beta| \cos (\theta-\phi))=0  \\
&\quad\ \text { and }
|\alpha|(e \cos (\Delta) \cos (\theta)-\sin (\Delta) \sin (\theta))+f|\beta| \cos (\Delta) \cos (\phi)=0.
\end{align*}The result of this lemma is given by this condition.

 \end{proof}
 
\end{lemma}
 We focus on the model satisfying $\theta=0,e=0,f=-1,\phi=-\frac{\pi}{2}\ \text{and}\ |\alpha|,|\beta|=\frac{1}{\sqrt{2}}$ in the next section. These conditions satisfy the conditions of Lemma 2. Here, $M_R, M_I,$ and $\tilde{M}$ become
\[M_{R}=\frac{e^{i \Delta}}{\sqrt{2}}\left(\begin{array}{cc}
1 & -i \\
-i & 1
\end{array}\right), \quad M_{I}=\frac{e^{i \Delta}}{\sqrt{2}}\left(\begin{array}{cc}
i & -1 \\
-1 & i
\end{array}\right),\]

\begin{align}
\tilde{M}=\frac{1}{\sqrt{2}}\left(\begin{array}{cccc}
\cos \Delta & \sin \Delta & -\cos \Delta & \sin \Delta \\
\sin \Delta & \cos \Delta & \sin \Delta & -\cos \Delta \\
\sin \Delta & -\cos \Delta & -\sin \Delta & -\cos \Delta \\
-\cos \Delta & \sin \Delta & -\cos \Delta & -\sin \Delta
\end{array}\right).
\end{align}
Note that when $\Delta=\frac{3}{4}\pi$, $\tilde{M}$ is nothing but  the Grover matrix. 

\section{Analysis on DQWs via 4-state LQWs}\label{main}

Here we introduce the method of Grimmett, Janson, and Scudo \cite{GJS}. We apply this method to derive the weak limit theorem of a one-parameter family of the 4-state Grover walk generated by DQWs in the last part. 
Since $\hat{U}(k)$ is a unitary matrix, it has eigenvalues $\lambda_{j}\  (j=1,2,3,4)$ whose absolute values are 1 and also has the orthonormal eigenvectors $\ket{v_j}$ associated with $\lambda_{j}.$ Using the spectral decomposition of $\hat{U}(k)$,
we have
\[\hat{U}(k)=\sum_{j=1}^{4} \lambda_{j}\left|v_{j}\right\rangle\left\langle v_{j}\right|,
\quad
\hat{U}(k)^{n}=\sum_{j=1}^{4} \lambda_{j}^{n}\left|v_{j}\right\rangle\left\langle v_{j}\right|.\]Putting $D=i\frac{d}{dk}$, the $r$-th moment of $X_n$ can be expressed as

\[E(X_n^r)=\int_{-\pi}^{\pi} \hat{\Psi}^*_n(k) D^r \hat{\Psi}_n(k)\frac{dk}{2\pi},\]
where $\hat{\Psi}^{*}={}^T\overline{\hat{\Psi}}(k)$ is the hermitian conjugate of $\hat{\Psi}(k)$. Furthermore,
\[D^r\hat{\Psi}_n(k)=\displaystyle\sum_{j=1}^4(n)_r\lambda^{n-r}_j({D\lambda_j})^r\bra{v_j}\hat{\Psi}_0(k)\ket{v_j} +O(n^{r-1}).\]Here $(n)_r=n(n-1)(n-2)\cdots(n-r+1)$, and $O(f(n))$ satisfies $\displaystyle\limsup_{n\rightarrow \infty}|O(f(n))/f(n)|\leq C$ for a finite constant $C\geq0$.
Since $\hat{\Psi}_n^*(k)=\sum_{j=1}^4 \lambda^{-1}_j\Psi_0^*(k)\ket{v_j}\bra{v_j}$, the $r$-th moment of $X_n$ becomes
\[E(X^r_n)=\int_{-\pi}^\pi\displaystyle\sum_{j=1}^4(n)_r\left(\frac{D\lambda_j}{\lambda_j}\right)^r|\bra{v_j}\hat{\Psi}_0(k)|^2 \frac{dk}{2\pi}+O(n^{r-1}).\]
Then, the limit of the $r$-th moment of $X_n/n$ is obtained as
\begin{equation}\label{limit}\lim_{n\rightarrow \infty}E\left(\left(\frac{X_n}{n}\right)^r\right)=\int_{-\pi}^\pi\displaystyle\sum_{j=1}^4\left(\frac{D\lambda_j}{\lambda_j}\right)^r|\bra{v_j}\hat{\Psi}_0(k)|^2 \frac{dk}{2\pi}.\end{equation}

In Theorem \ref{main theorem}, for the generalization, let the initial state at the origin be an arbitrary 4-state complex vector. However, note that only when it is a real vector, there is a corresponding DQW:
\[(\iota\Psi_{0})(x)=
\left(\begin{array}{l}
\phi_{1} \\
\phi_{2}  \\
\phi_{3}  \\
\phi_{4}  
\end{array}\right)\delta_0(x)\in \mathbb{C}^4.\]

\begin{theorem}
\label{main theorem}
$\\\mathrm{(A)}$When $ \Delta \neq \pm\frac{\pi}{4},\pm \frac{3\pi}{4}$, we have
\[\lim_{n\rightarrow \infty}\mathbb{P}\left(\frac{X_n}{n}\leq x\right)=\int_{-\infty}^x \{A\delta_0(y)+f(y)I_{(-|p|,|p|)}(y)\}dy.\]
Here $\delta_0(x)$ is the Dirac delta function at the origin, and
\[A=1-\frac{d_0}{2}-\frac{1-\sqrt{1-p^2}}{2p^2}d_2,\]
\[f(x)=\frac{\sqrt{1-p^2}}{2\pi p^2\sqrt{p^2-x^2}(1-x^2)}(p^2d_0+pd_1x+d_2x^2),\]
where
\begin{align*}
d_{0}&=\left|\phi_{1}-\phi_{2}\right|^{2}+\left|\phi_{3}-\phi_{4}\right|^{2}, 
\\
d_{1}&=p\left(\left|\phi_{2}-\phi_{4}\right|^{2}-\left|\phi_{1}-\phi_{3}\right|^{2}\right)+q\left(\left|\phi_{2}+\phi_{3}\right|^{2}-\left|\phi_{1}+\phi_{4}\right|^{2}\right) ,
\\
d_{2}&=p q\left(\left|\phi_{1}+\phi_{2}\right|^{2}-\left|\phi_{3}+\phi_{4}\right|^{2}\right)+2 p^{2} \mathfrak{R}\left(\overline{\left(\phi_{1}-\phi_{4}\right)}\left(\phi_{2}-\phi_{3}\right)\right)+2 q^{2} \Re\left(\overline{\left(\phi_{1}+\phi_{3}\right)}\left(\phi_{2}+\phi_{4}\right)\right),
\end{align*}
and
\[p=\frac{1}{\sqrt{2}}(\cos \Delta -\sin \Delta),
\quad
q=\frac{1}{\sqrt{2}}(\cos \Delta +\sin \Delta).\] 
 $\mathrm{(B)}$ When $\Delta = \frac{\pi}{4}\ \text{or}\ \Delta=-\frac{3\pi}{4}$, for $x\in\mathbb{Z}$ and $m\in\mathbb{Z}_{\geq}=\{0,1,2,\cdots\}$, the quantum state $ (\tilde{U}^n\iota\Psi)(x)$ can be expressed as follows:
 
 \noindent$\mathbf{case\ 1}$ $n=4m:$
\[ (\tilde{U}^n\iota\Psi)(x)=\left(\begin{array}{ c }
\phi _{1}\\
\phi _{2}\\
\phi _{3}\\
\phi _{4}
\end{array}\right) \delta _{0}(x),\]
$\mathbf{case\ 2}$ $n=4m+1:$
\[ (\tilde{U}^n\iota\Psi)(x)=\frac{( -1)^{j}}{2}\left(\begin{array}{ c }
\phi _{1} +\phi _{2} -\phi _{3} +\phi _{4}\\
0\\
\phi _{1} -\phi _{2} -\phi _{3} -\phi _{4}\\
0
\end{array}\right) \delta _{-1}(x) +\frac{( -1)^{j}}{2}\left(\begin{array}{ c }
0\\
\phi _{1} +\phi _{2} +\phi _{3} -\phi _{4}\\
0\\
-\phi _{1} +\phi _{2} -\phi _{3} -\phi _{4}
\end{array}\right) \delta _{1}(x),\]$\mathbf{case\ 3}$ $n=4m+2:$
\[ (\tilde{U}^n\iota\Psi)(x)=\frac{1}{2}\left(\begin{array}{ c }
\phi _{2} +\phi _{4}\\
0\\
\phi _{2} +\phi _{4}\\
0
\end{array}\right) \delta _{-2} (x)+\frac{1}{2}\left(\begin{array}{ c }
\phi _{2} -\phi _{4}\\
\phi _{1} -\phi _{3}\\
\phi _{4} -\phi _{2}\\
\phi _{3} -\phi _{1}
\end{array}\right) \delta _{0} (x)+\frac{1}{2}\left(\begin{array}{ c }
0\\
\phi _{1} +\phi _{_{3}}\\
0\\
\phi _{1} +\phi _{3}
\end{array}\right) \delta _{2}(x),\]$\mathbf{case\ 4}$ $n=4m+3:$
\[ (\tilde{U}^n\iota\Psi)(x)=\frac{( -1)^{j}}{2}\left(\begin{array}{ c }
\phi _{2} -\phi _{4}\\
\phi _{2} +\phi _{4}\\
\phi _{2} -\phi _{4}\\
-\phi _{2} -\phi _{4}
\end{array}\right) \delta _{-1}(x) +\frac{( -1)^{j}}{2}\left(\begin{array}{ c }
\phi _{1} +\phi _{3}\\
\phi _{1} -\phi _{3}\\
-\phi _{1} -\phi _{3}\\
\phi _{1} -\phi _{3}
\end{array}\right) \delta _{1}(x),\]where $j=0\ (\Delta=\frac{\pi}{4}),\  j=1\ (\Delta=\frac{-3\pi}{4})$, and  $\delta_{x_0}(x)$ is a function defined by ($\ref{Delta function}).$

\noindent$\mathrm{(C)}$ When $\Delta = \frac{3\pi}{4}\ \text{or}\ \Delta=-\frac{\pi}{4}$, for $x\in\mathbb{Z}$ and $m\in\mathbb{Z}_{\geq}$, the quantum state $ (\tilde{U}^n\iota\Psi)(x)$  can be expressed as follows:

\noindent$\mathbf{case\ 1}$ $n=2m:$
\[ (\tilde{U}^n\iota\Psi)(x)=\left(\begin{array}{ c }
\phi _{1} -\phi _{3}\\
0\\
\phi _{3} -\phi _{1}\\
0
\end{array}\right)\frac{\delta _{-n}(x)}{2} +\left(\begin{array}{ c }
\phi _{1} +\phi _{3}\\
\phi _{2} +\phi _{4}\\
\phi _{1} +\phi _{3}\\
\phi _{2} +\phi _{4}
\end{array}\right)\frac{\delta _{0}(x)}{2} +\left(\begin{array}{ c}
0\\
\phi _{2} -\phi _{4}\\
0\\
\phi _{4} -\phi _{2}
\end{array}\right)\frac{\delta _{n}(x)}{2},\]
\noindent$\mathbf{case\ 2}$ $n=2m+1:$
\begin{align*}
 (\tilde{U}^n\iota\Psi)(x)=( -1)^{j}\left(\left(\begin{array}{ c}
\phi _{1} -\phi _{3}\\
0\\
\phi _{3} -\phi _{1}\\
0
\end{array}\right)\right.
&\frac{\delta _{-n}(x)}{2} 
-\left(\begin{array}{ c }
\phi _{2} +\phi _{4}\\
0\\
\phi _{2} +\phi _{4}\\
0
\end{array}\right)\frac{\delta _{-1}(x)}{2} 
\\
&-\left(\begin{array}{ c }
0\\
\phi _{1} +\phi _{3}\\
0\\
\phi _{1} +\phi _{3}
\end{array}\right)\left.\frac{\delta _{1}(x)}{2} +\left(\begin{array}{ c }
0\\
\phi _{2} -\phi _{4}\\
0\\
\phi _{4} -\phi _{2}
\end{array}\right)\frac{\delta _{n}(x)}{2}\right),
\end{align*}
where $j=0\ (\Delta=-\frac{\pi}{4}),\  j=1\ (\Delta=\frac{3\pi}{4})$.

\vspace{5mm}
\textup{We should remark that the limit measure has a limit density function with finite support  $(-|p|,|p|)$ and the Dirac delta function at the origin. Moreover, the Dirac delta part contributes to localization of the model. The result of (A) is given under the condition of $ \Delta \neq \pm\frac{\pi}{4},\pm \frac{3\pi}{4}$. However, with respect to the case $ \Delta = \pm\frac{\pi}{4},\pm \frac{3\pi}{4}$, we can get the quantum states for arbitrary $n$ by calculating the inverse Fourier transform of $\hat{\Psi}_n(k)$. In particular, when $\Delta=\frac{\pi}{4},-\frac{3}{4}\pi$, the period of quantum states is 4, since  $\hat{U}(k)^4$ is the identity matrix.} 

 \begin{proof}
We start with eigenvalues of $\hat{U}(k)$ which  are given by
\begin{align*}\lambda_1=1,\quad \lambda_2&=-1,\quad \lambda_3=p\cos k+i\sqrt{1-p^2\cos ^2 k},\quad \lambda_4=p\cos k-i\sqrt{1-p^2\cos ^2 k}.\end{align*}
  Under the assumptions of $\Delta \neq \frac{3\pi}{4}\ \text{and}\ \Delta\neq-\frac{\pi}{4}$, the normalized eigenvectors corresponding to $\lambda_j(k)\ (j=3,4)$ are obtained with appropriate normalization factors $N_j$ as below:
 \begin{align*} \ket{v_j(k)}=N_j
\begin{pmatrix}
x_1(k,j)x_2(k,j)x_3(k,j) \\ \overline{x_1}(k,j)x_3(k,j)x_4(k,j)\\ x_1(k,j)x_2(k,j)\overline{x_4}(k,j) \\ x_2(k,j)x_3(k,j)x_4(k,j)
\end{pmatrix},
\end{align*}
where
\begin{align*}
&x_1(k,j)=\lambda_j e^{ik}+\sqrt{2}\sin{\Delta},
\quad x_2(k,j)=\lambda^{-1}_j e^{-ik}-\sqrt{2}\cos{\Delta},\\
&x_3(k,j)=\lambda^{-1}_j e^{ik}+\sqrt{2}\sin{\Delta},
\quad x_4(k,j)=\lambda_j e^{-ik}-\sqrt{2}\cos{\Delta}.
\end{align*}Additionally, 
\[\frac{D\lambda_3}{\lambda_3}=\frac{-p\sin{k}}{\sqrt{1-p^2\cos {k}^2}},\quad \frac{D\lambda_4}{\lambda_4}=\frac{p\sin{k}}{\sqrt{1-p^2\cos {k}^2}}.\] After a direct calculation of (\ref{limit}) by putting $D\lambda_j/\lambda_j=x$ for $j=3,4$, and assuming $\Delta \neq \frac{\pi}{4}\ \text{and}\ \Delta\neq-\frac{3\pi}{4}$, we get the continuous part $f(x)$ of the limit measure for case (A). Moreover, the coefficient of the Dirac delta function denoted by $A$, is obtained in the following way:
\begin{align*}A&=\int_{-\pi}^{\pi}p_1(k)\frac{dk}{2\pi}+\int_{-\pi}^{\pi}p_2(k)\frac{dk}{2\pi}\\
&=1-\left(\int_{-\pi}^{\pi}p_{3}(k)\frac{dk}{2\pi}+\int_{-\pi}^{\pi}p_{4}(k)\frac{dk}{2\pi}\right)\\
&=1-\left(\int_{-|p|}^{|p|}\frac{\sqrt{1-p^2}d_0}{2\pi \sqrt{p^2-x^2}(1-x^2)}dx+\int_{-|p|}^{|p|}\frac{\sqrt{1-p^2}d_2x^2}{2\pi p^2\sqrt{p^2-x^2}(1-x^2)}dx\right)\\
&=1-\frac{d_0}{2}-\frac{1-\sqrt{1-p^2}}{2p^2}d_2.
\end{align*}
In the case of $\Delta = \frac{\pi}{4}\ \text{and}\ \Delta=-\frac{3\pi}{4}$, we have 
\[\hat{U}(k)^n=\begin{cases}
E, & n=4m,\\
\displaystyle
\frac{( -1)^{j}}{2}\left(\begin{array}{ c c c c }
e^{ik} & e^{ik} & -e^{ik} & e^{ik}\\
e^{-ik} & e^{-ik} & e^{-ik} & -e^{-ik}\\
e^{ik} & -e^{ik} & -e^{ik} & -e^{ik}\\
-e^{-ik} & e^{-ik} & -e^{-ik} & -e^{-ik}
\end{array}\right) , & n=4m+1,\\
\displaystyle
\frac{1}{2}\left(\begin{array}{ c c c c }
0 & 1+e^{2ik} & 0 & -1+e^{2ik}\\
1+e^{-2ik} & 0 & -1+e^{-2ik} & 0\\
0 & -1+e^{2ik} & 0 & 1+e^{2ik}\\
-1+e^{-2ik} & 0 & 1+e^{-2ik} & 0
\end{array}\right) , & n=4m+2,\\
\displaystyle
\frac{( -1)^{j}}{2}\left(\begin{array}{ c c c c }
e^{-ik} & e^{ik} & e^{-ik} & -e^{ik}\\
e^{-ik} & e^{ik} & -e^{-ik} & e^{ik}\\
-e^{-ik} & e^{ik} & -e^{-ik} & -e^{ik}\\
e^{-ik} & -e^{ik} & -e^{-ik} & -e^{ik}
\end{array}\right) , & n=4m+3,
\end{cases}\]
where $m\in\mathbb{Z}_{\geq}$, $j=0\ (\Delta = \frac{\pi}{4})$, $j=1\ (\Delta = -\frac{3\pi}{4})$ and $E$ is an identity matrix. By the inverse Fourier transform of $\hat{U}(k)^n\begin{pmatrix}
\phi_1 & \phi_2 & \phi_3 & \phi_4
\end{pmatrix}^{\top}$, we get the result of (B).
In the case of $\Delta = \frac{3\pi}{4}\ \text{and}\ \Delta=-\frac{\pi}{4}$, from the spectral decomposition of $\hat{U}(k)$, we have 
\[
\begin{aligned}
\hat{\Psi}_n(k)=(-1)^{jn}\left(\frac{1^{n}}{4}\left(\begin{array}{ c c c c }
1 & -e^{ik} & 1 & -e^{ik}\\
-e^{-ik} & 1 & -e^{-ik} & 1\\
1 & -e^{ik} & 1 & -e^{ik}\\
-e^{-ik} & 1 & -e^{-ik} & 1
\end{array}\right)\right. +\frac{( -1)^{n}}{4}\left(\begin{array}{ c c c c }
1 & e^{ik} & 1 & e^{ik}\\
e^{-ik} & 1 & e^{-ik} & 1\\
1 & e^{ik} & 1 & e^{ik}\\
e^{-ik} & 1 & e^{-ik} & 1
\end{array}\right) & \\
+\frac{e^{ink}}{2}\left(\begin{array}{ c c c c }
1 & 0 & -1 & 0\\
0 & 0 & 0 & 0\\
-1 & 0 & 1 & 0\\
0 & 0 & 0 & 0
\end{array}\right) +\left.\frac{e^{-ink}}{2}\left(\begin{array}{ c c c c }
0 & 0 & 0 & 0\\
0 & 1 & 0 & -1\\
0 & 0 & 0 & 0\\
0 & -1 & 0 & 1
\end{array}\right)\right) \left(\begin{array}{ c }
\phi _{1}\\
\phi _{2}\\
\phi _{3}\\
\phi _{4}
\end{array}\right)& 
\end{aligned}\]
where $j=0\ (\Delta = -\frac{\pi}{4})$ and $j=1\ (\Delta = \frac{3\pi}{4}).$ By the inverse Fourier transform of $\hat{\Psi}_n(k)$, we get the result of (C).
 \end{proof}

 \end{theorem}
 In order to demonstrate the validity of our results in Theorem \ref{main theorem}, we show comparisons between rescaled probability distributions at time $n=100$ and continuous parts of the limit measures with numerical simulations in Figs. \ref{simu1 fig} and \ref{simu2 fig}. For these cases, since $d_1=0$, the limit density functions are symmetric. In Fig. \ref{simu1 fig}, the limit density functions are  restricted in $|x|<|p|=\frac{1}{\sqrt{2}} \fallingdotseq 0.7071$ with  $\lim_{x\rightarrow \pm|p|}f(x)=\infty $, which means that they show peaks at the leftmost and rightmost. In the case of Fig. \ref{simu1 fig}-(a), the limit measure has the Dirac delta function with a positive coefficient, and the probability distribution shows the peak around the origin. With respect to Fig. \ref{simu1 fig}-(b), the probability distribution does not show the peak around the origin, since the coefficient of the Dirac delta function is $0$. In Fig. \ref{simu2 fig}, $\lim_{x\rightarrow \pm|p|}f(x)=0 $ holds, which means that their probability distributions do not show the peak at the leftmost and the rightmost. The limit density function in Fig. \ref{simu2 fig}-(a) shows a convex shape, and with respect to Fig. \ref{simu2 fig}-(b), $p \fallingdotseq 0.966$, and the limit density function shows slight concavity.

\begin{figure}[H]

\begin{subfigure}{0.5\textwidth}
\centering

\includegraphics[width=1\linewidth, height=4.5cm]{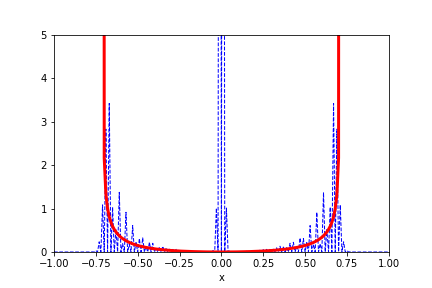} 
\caption{$\Delta =\frac{\pi }{2},\ \phi _{1}=\phi _{2}=\phi _{3}=\phi _{4} =\frac{1}{2} $}
\end{subfigure}
\begin{subfigure}{0.5\textwidth}
\centering

\includegraphics[width=1\linewidth, height=4.5cm]{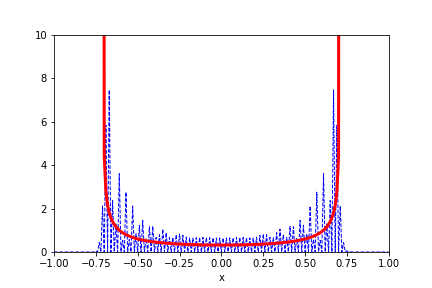}
\caption{\footnotesize $\Delta =\frac{\pi}{2},\  \phi _{1}=\phi _{4}=\frac{1}{2},\ \phi _{2}=\phi _{3} =-\frac{1}{2} $}

\end{subfigure}

\addtocounter{figure}{-1} 
\fcaption{The comparisons between rescaled probability distributions at time $n=100$ and continuous parts of the limit measures satisfying $\lim_{x\rightarrow \pm|p|}f(x)=\infty $.  }
\small{
}
\label{simu1 fig}
\end{figure}

\begin{figure}[H]
\begin{subfigure}{0.5\textwidth}
\centering
\includegraphics[width=1\linewidth, height=4.5cm]{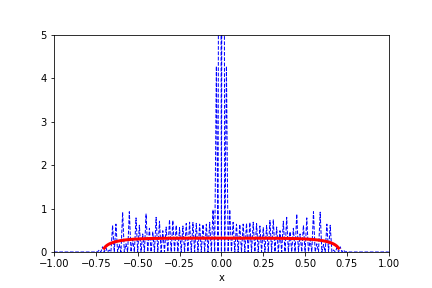} 
\caption{$\Delta =\frac{\pi }{2},\ \phi _{1}=\phi _{3}=\frac{1}{2},\ \phi _{2}=\phi _{4} =-\frac{1}{2} $}
\end{subfigure}
\begin{subfigure}{0.5\textwidth}
\centering
\includegraphics[width=1\linewidth, height=4.5cm]{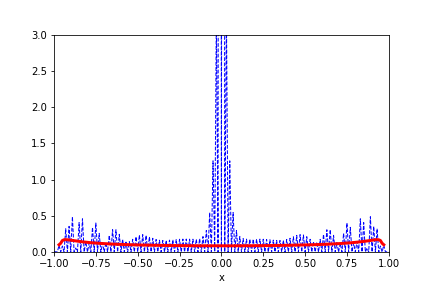}
\caption{\footnotesize $\Delta =-\frac{\pi}{6} \pi,\  \phi _{1}=\phi _{3}=\frac{1}{2},\ \phi _{2}=\phi _{4} =-\frac{1}{2} $}

\end{subfigure}
\addtocounter{figure}{-1} 
\fcaption{The comparisons between rescaled probability distributions at time $n=100$ and continuous parts of the limit measures satisfying $\lim_{x\rightarrow \pm|p|}f(x)=0 $. }
\small{
}
\label{simu2 fig}
\end{figure}

\section{Summary}
In this paper, we introduced 2-state DQWs on a line as the extensions of 2-state LQWs. The time evolution of the DQW was defined using two matrices, $M_R$ and $M_I$.  In Lemma 1, we proved that to preserve a sum of the probability measures, $M_R$ and $M_I$ must be unitary matrices, and $M_R^*M_I$ must be a real matrix. Additionally, it was revealed that 2-state DQWs can be regarded as a class of 4-state LQWs, which includes the Grover walk. In Lemma 2, we showed the necessary and sufficient condition for 1 and -1 to be eigenvalues of the time evolution operator of 4-state LQWs corresponding to DQWs in the Fourier space. This lemma helps us construct the generalized 4-state Grover walk, and we focused on a specific one-parameter family of the 4-state Grover walk. Then, for this model, Theorem 2 reveals asymptotic behaviors by deriving the limit measures. Moreover, we showed that localization and the spreading phenomenon can occur simultaneously in DQWs, and there are two types of continuous parts of the limit measures. They can be visually confirmed by computer simulations in Figs. \ref{simu1 fig} and \ref{simu2 fig}.

\section{Acknowledgments}
The author expresses sincere thanks and gratitude to Norio Konno for helpful comments and discussion. The author would also like to thank Kei Saito for useful advice.

\nonumsection{References}

\end{document}